\documentclass[lettersize,journal]{IEEEtran}
\usepackage[dvipsnames]{xcolor}
\usepackage{amsmath,amsfonts}
\usepackage{algorithmic}
\usepackage{array}
\usepackage[caption=false,font=normalsize,labelfont=sf,textfont=sf]{subfig}
\usepackage{textcomp}
\usepackage{stfloats}
\usepackage{url}
\usepackage{verbatim}
\usepackage{graphicx}
\def\BibTeX{{\rm B\kern-.05em{\sc i\kern-.025em b}\kern-.08em
    T\kern-.1667em\lower.7ex\hbox{E}\kern-.125emX}}
\usepackage{balance}
\usepackage[style=ieee, citestyle=numeric-comp, backend=biber]{biblatex}
\begin{document}
\title{Reformulated Fourier Modal Method with improved near field computations}
\author{Sergey Spiridonov, and Alexey A. Shcherbakov \\ Department of Physics, ITMO University, St-Petersburg, Russia}


\maketitle

\begin{abstract}
In this paper we propose a new formulation of the Fourier Modal Method based on an alternative treatment of interface conditions allowing us to overcome the effect of the Gibbs phenomenon. Explicit consideration of the interface conditions for the discontinuous part of the field leads to an equation for the eigenvalue problem, which can be written in an inversion-free form. The results of the method are in good agreement with the results for the classical approach based on the Li factorization rules both for dielectric and metallic gratings. Moreover, the developed method allows calculating the near field much more accurately, and may find its applications in sensing and nonlinear optics.
\end{abstract}


\section{Introduction}
\IEEEPARstart{T}{he} Fourier Modal Method (FMM) also known as the Rigorous Coupled-Wave Analysis (RCWA) \cite{FMM95} is a widely used tool for calculating linear interaction of electromagnetic radiation with periodic structures. It is based on the Fourier plane wave decomposition of all periodic functions and has gained popularity due to its relative simplicity and efficiency for a wide class of problems. This method is well-suited for analyzing periodic grating metal and dielectric structures, calculating their eigenmodes and high-Q resonances \cite{song2018first, gansch2016measurement, suzuki2019simulation}, solar cell simulations \cite{robertson2019efficient, solar, Tmatrix,effectiveMedia}, in particular, the method is suitable for supercell calculations \cite{supercellFirst}. In addition, the FMM can be also used to find plasmonic resonances \cite{plasmonic} and to solve non-linear optics problems \cite{nonlinear,secondHarmonic}.
 
 Knop, Moharam and Gaylord first formulated the method \cite{Knop, FMM81,FMM95,FMM2} and since then it has been developed for multidimensional and arbitrary shaped gratings \cite{Peng:95, Popov:00, Gushchin:10} and improved with using normal vector approach \cite{Popov:01, Davids2021} and adaptive spatial resolution \cite{Granet:99, Granet:01, Weiss:09, yala2009fourier, Essig:10}. One of the main problems of the FMM was poor convergence for TM polarization and metal gratings, implicitly caused by the well-known Gibbs phenomenon. Granet and Lalanne proposed a receipt to improve the rate of convergence \cite{Granet, Lalanne} and then Li formulated special rules for the factorization of products of discontinuous functions in the Fourier space \cite{li1996use, li1997new}. The approach substantially improves the convergence of the method, but does not completely eliminate the influence of the Gibbs phenomenon, so it evolved in different directions. For example, J. Li with co-authors proposed an alternative formulation, using the Cayley-Hamilton theorem and reducing the computational time of the method \cite{Li2020}. Sauvan and Edee developed alternative methods using orthogonal polynomials systems instead of plane-wave decomposition \cite{sauvan2004truncation, Edee:11}. This approach eliminates the Gibbs effect problem by a cost of partial loss of method's flexibility in terms of coupling multilayer structures and potential applications of fast numerical schemes (e.g., \cite{FFT, skobelev2013analysis}). In addition, the convergence of the near field, which is important for certain applications like sensing and non-linear optics, was outside the scope of the review for a long time. Lalanne and Weismann \cite{lalanne1998computation, weismann2015accurate} showed that the near field converged by an order of magnitude worse than the far field and proposed a way to calculate the near field more accurately using an appropriate order of multiplications by permittivity dependent functions.
 
 In this paper we propose an alternative formulation of FMM taking into account the boundary conditions for discontinuous part of electromagnetic field explicitly, thereby reducing the impact of the Gibbs phenomenon. We supplement the usual Fourier basis with additional jump functions, which amplitudes appear to be unambiguously defined by interface conditions, so that the numerical problem remains to be formulated for the Fourier amplitude vectors, as usual. Introduction of known step functions into Fourier space methods already demonstrated its profitability, see e.g. \cite{qing2001improved}. In order to test the idea we consider a simple problem of a photonic crystal slab being periodic in one dimension. Section \ref{sec:implementation} provides a formulation of our approach. Section \ref{sec:numerical} shows numerical results on the convergence of the reflection coefficient for the new method in comparison with the classical FMM, and also demonstrates the convergence of the near field. Section \ref{sec:conclusion} summarizes the results and shows that the new approach gives accurate values of the near field already in Fourier space.   

\section{New implementation}
\label{sec:implementation}

\noindent In this section we consider the main equations of the new method and compare then with the classical formulation with the Li factorization. The main idea is to take into account the boundary conditions for the discontinuous part of the electric field explicitly, attain new equations for the near field, and reduce the impact of the Gibbs effect.

Consider a simple case of a one-dimensional photonic crystal slab consisting of two materials with permittivities $\varepsilon_1$ and $\varepsilon_2$. The grating structure is surrounded by a substrate and superstrate having the dielectric permittivities $\varepsilon_{sub}$ and $\varepsilon_{sup}$ respectively. The structure is periodic along the $x$-direction with a period $\Lambda$, and a filling factor for the first material is $\alpha$. The structure is invariant in the $y$-direction, while $z$-axis is perpendicular to the grating boundaries. We consider the TM polarization only since the TE case is not modified for non-magnetic media. The Maxwell's equations for such a problem read:
\begin{equation}\label{eq:maxw_h}
\begin{aligned}
\frac{\partial H_{y}}{\partial z} =i \omega \varepsilon(x) E_{x}, \\
\frac{\partial H_{y}}{\partial x} =-i \omega \varepsilon(x) E_{z}, \\
\frac{\partial E_{x}}{\partial z}-\frac{\partial E_{z}}{\partial x} =i \omega \mu_0 H_{y},
\end{aligned}
\end{equation}
where $\varepsilon(x) = \varepsilon_1$ for $n\Lambda\leq x<(n+\alpha)\Lambda$, $n\in\mathbb{Z}$, and $\varepsilon(x) = \varepsilon_2$ otherwise. $E_i, H_i$ are the electric and magnetic field components respectively, $\mu_0$ is the vacuum magnetic permeability. The Floquet-Bloch theorem for the field $E_x(x,z)$ yields
\begin{equation}\label{Bloch}
E_x\left(x,z\right)=\exp\left(ik_0x\right)e_x\left(x,z\right)
\end{equation}
where $k_0$ is the Bloch wavenumber, and $e_x$ is a purely periodic function of the coordinate $x$.

 Instead of developing $e_x$ into its Fourier series as is done within the FMM, we consider the singularities of the field separately in the coordinate space. The field component $E_x$ is a discontinuous function of the coordinate $x$. We represent it inside the grating structure as a superposition of a continuous and discontinuous parts:
\begin{equation}\label{Ex_cd}
    E_x(x,z) = E^{c}_x(x,z) + E^{d}_x(x,z).
\end{equation}
This decomposition is not unique. To specify the second component consider a sum of some known periodic linear piece-wise continuous functions $g_k(x)$ each having a single point of discontinuity $x_k$, $k=1,\dots,N_d$, coinciding with a corresponding discontinuity point of the permittivity function $\varepsilon(x)$. Here $N_d = 2$ is the number of such points for the photonic crystal slab under consideration. The decomposition reads
\begin{equation}\label{Ed_gk}
    E_{x}^{d}= \exp\left(ik_0x\right)\sum_{k=1}^{N_d} \xi_k(z) g_{k}(x)
\end{equation}
Also, let us take an additional assumption $g_k(x_k+0) - g_k(x_k-0) = 1$ and $g_k(x_k+0) = - g_k(x_k-0) = 1/2$. Note that $g_k(x_j+0) = g_k(x_j-0)$, $j\neq k$, by definition.

The interface field continuity condition at $k$-th point of discontinuity $x_k$ is
\begin{equation}\label{xk_continuity}
    \varepsilon_{k}E_{x}\left(x_{k}-0\right)=\varepsilon_{k+1}E_{x}\left(x_{k}+0\right),
\end{equation}
Substitution of Eqs. (\ref{Ex_cd}) and (\ref{Ed_gk}) into the latter relation gives
\begin{multline}\label{xi_eq}
   \xi_k(z) + 2\dfrac{\varepsilon_{k+1}-\varepsilon_{k}}{\varepsilon_{k}+\varepsilon_{k+1}}\sum_{q=1,q\neq k}^{N_d} \xi_q(z) g_q(x_k) = \\ = -2\exp\left(-ik_0 x_k\right)\dfrac{\varepsilon_{k+1}-\varepsilon_{k}}{\varepsilon_{k}+\varepsilon_{k+1}} E_x^c (x_k,z).
\end{multline}
The number of equations in (\ref{xi_eq}) equals to the number of discontinuity points $N_d$, which is 2 in our case. The index $k+1$ should be treated within a modulo $N_d$ sum.

Bloch-Fourier decomposition of the continuous part of the field $E^c_x$ allows to relate the discontinuous field amplitudes with corresponding Fourier amplitudes as follows:
\begin{equation}\label{Exd}
    \xi_k(z) = \sum_{q=1}^{N_d} G_{kq} \sum_{m\in\mathbb{Z}} \exp(ik_{xm}x_{q}) E_{xm}^{c}(z),
\end{equation}
where $k_{xm} = k_0+2\pi m/\Lambda$ and $E_{xm}^c$ is the Fouier amplitude of the continuous field. The constant matrix $G$ depends only on structure parameters and predefined functions $g_k$, and writes
\begin{multline}\label{eq:G}
    G =  \exp\left(-ik_0 x_k\right) \times \\ \times \left(\left\{ \dfrac{1}{2}\dfrac{\varepsilon_k+\varepsilon_{k+1}}{\varepsilon_{k}-\varepsilon_{k+1}}\delta_{kq} - (1-\delta_{kq}) g_q(x_k)  \right\}_{k,q=1}^{N_d} \right)^{-1},
    \end{multline}
with $\delta_{kq}$ being the Kronecker symbol.

The developed equations allow rewriting the $E_x$ and the product $\varepsilon E_x$ in right-hand part of the Maxwell's equations (\ref{eq:maxw_h}) via the continuous field part $E_x^c$ only. Denote vectors of the Fourier amplitudes of these functions as $[E_x]$, $[\varepsilon E_{x}]$, and $[E_x^c]$ respectively. Then, Eqs.(\ref{Ex_cd})-(\ref{eq:G}) yield
\begin{equation}\label{eq:ex_exc}
    [E_x]= [E_x^c] + P [E_x^c]
\end{equation}
\begin{equation}\label{eq:ex_eps_exc}
    [\varepsilon E_{x}] = \left( [[\varepsilon]]  + Q \right) [E_x^c].
\end{equation}
Here $[[\varepsilon]]$ is the Toeplitz matrix of permittivity Fourier components. Matrices $P$ and $Q$ are derived from Eq.~(\ref{Exd}) and can be calculated analytically as well as $[[\varepsilon]]$. Explicitly they are given in the Appendix, Eqs.~(\ref{eq:P-matrix}, \ref{eq:Q-matrix}). Introduction of $Q$ and $P$ takes into account the interface conditions for the discontinuous field component.

Having derived the Fourier vectors corresponding to the discontinuous field and its product with the permittivity, the Fourier transform of the Maxwell's equations (\ref{eq:maxw_h}) becomes
\begin{equation} \label{eq:maxw_fourier}
    \begin{gathered}
        \dfrac{\partial}{\partial z}[H_y] = i\omega ([[ \varepsilon]] + Q) [E_x^c] \\
        K_{x} [H_y] = -\omega [[\varepsilon]] [E_z] \\
        \dfrac{\partial}{\partial z}\left(\mathbb{I}+\mathcal{P}\right) [E_{x}^{c}] - iK_{x}[E_{z}]=i\omega\mu_{0}[H_y].
    \end{gathered}
\end{equation}
where $K_x = \mathrm{diag}\{k_{xm}\}$. Owing to the translational invariance of the photonic crystal, $z$-dependence of the fields is described by exponential propagators (as $[E_x^c] = \boldsymbol{e}_x^c\exp{\left(i\beta z\right)}$ and similarly for other field components), and Eq.~(\ref{eq:maxw_fourier}) reduces to an algebraic eigenvalue problem, which can be written in an inversion-free form:
\begin{equation}
\begin{gathered}
        \beta^{2}[[\varepsilon]]K_{x}^{-1}\left(\mathbb{I}+\mathcal{P}\right)\boldsymbol{e}_{x}^{c} = \\ = \left(\omega^{2}\mu_{0}[[\varepsilon]]K_{x}^{-1} -  K_{x}\right)\left([[\varepsilon]] +Q\right)\boldsymbol{e}_{x}^{c}
\end{gathered}
\end{equation}
With the latter equation being solved, the amplitude vector $h_y$ is attained as
\begin{equation} \label{eq:hy}
    \boldsymbol{h}_{y} = \omega B^{-1} \left( [[\varepsilon]] + Q\right) \boldsymbol{e}_{x}^{c}.
\end{equation}
with $B = \mathrm{diag} \{\beta_m\}$.

The difference between the new formulation and the approach implementing the Li’s factorization rules is as follows. When factoring a product of two discontinuous functions into a Fourier series Li's rules replace the matrix $[\varepsilon]$ with the matrix $[1/\varepsilon]^{-1}$. The alternative way suggests to keep the matrix $[\varepsilon]$ by adding the matrix $Q$ that takes into account the interface conditions explicitly and treats separately the continuous and discontinuous parts of the field.

The next steps are exactly the same as in the conventional FMM: construction of a T-matrix for an interface between the photonic crystal and a homogeneous medium, and derivation of a slab S-matrix based on the T-matrices of the two photonic crystal slab boundaries \cite{FMM95}. The only thing to account for is the relation (\ref{eq:ex_exc}) which defines the total near field.

\section{Numerical validation}
\label{sec:numerical}
\begin{figure}[!t]
\centering
\includegraphics[width=0.9\linewidth]{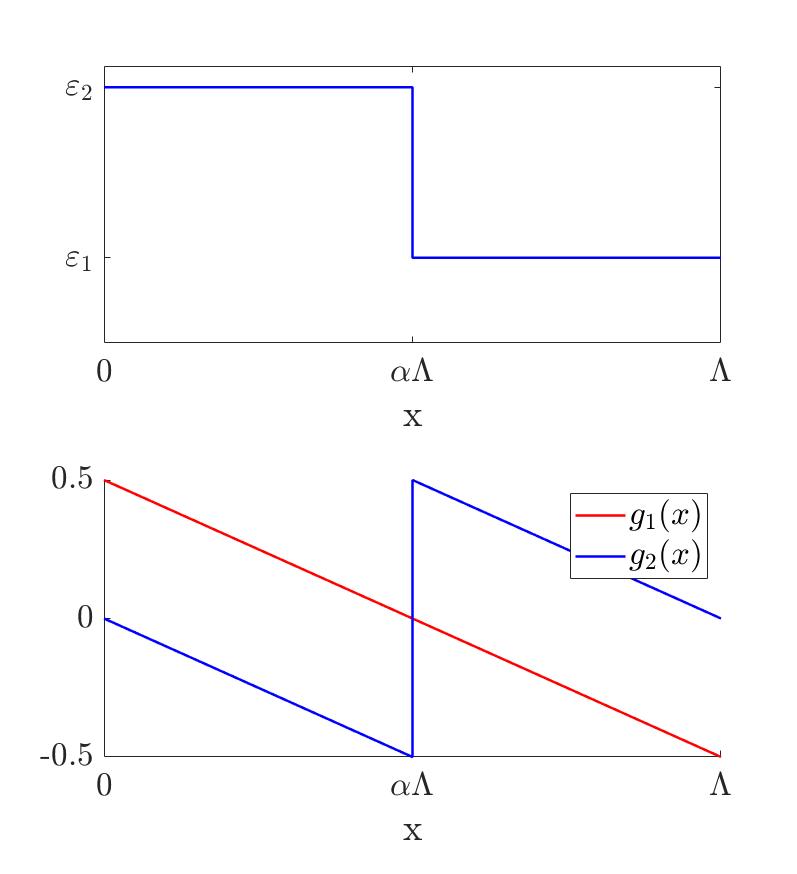}
\caption{Dielectric permittivity function of the example photonic crystal slab (above) and discontinuous functions given by Eq. (A.1), which are used in the
representation of the discontinuous field part, Eq. (4) (below).}
\label{fig:epsf}
\end{figure}

\noindent To validate the efficiency of the new approach we take two example structures and compare our method with the conventional FMM. Let us consider a slab of a dielectric one-dimensional photonic crystal, periodic in one dimension with the parameters $\varepsilon_1 = 1$, $\varepsilon_2 = 3.4^2$. The dielectric permittivity function is shown in Fig.~\ref{fig:epsf}. The dielectric permittivities of the substrate and superstrate media are $\varepsilon_{sub} = 1.45^2, \varepsilon_{sup} = 1$. The grating period is $\Lambda = 1\;\mu m$, the slab thickness is $h = 0.25\;\mu m$, the filling factor is $\alpha = 0.55$. The wavelength of the incident TM polarized plane wave is $\lambda = 0.51\;\mu m$. The angle of incidence is 1\textdegree. Problem setting of this kind is common within the modern dielectric nanophotonics which deals with high-Q resonances related to the bound states in the continuum \cite{Koshelev2020}. We also consider an example of  Au-grating with the parameters $\lambda = 0.51\;\mu m$, $\varepsilon_1 = 1$, $\varepsilon_2 = -2.5676 + 3.6391\mathrm{i}$, $\varepsilon_{sub} = \varepsilon_{sup} = 1$, $\Lambda = 1.15\;\mu m$, $\alpha = 0.55$, $h = 0.2\;\mu m$, because the case of metal gratings is more challenging \cite{li1996use}.

\begin{figure}[!t]
\begin{minipage}[h]{0.86\linewidth}
\center{\includegraphics[width=1\linewidth]{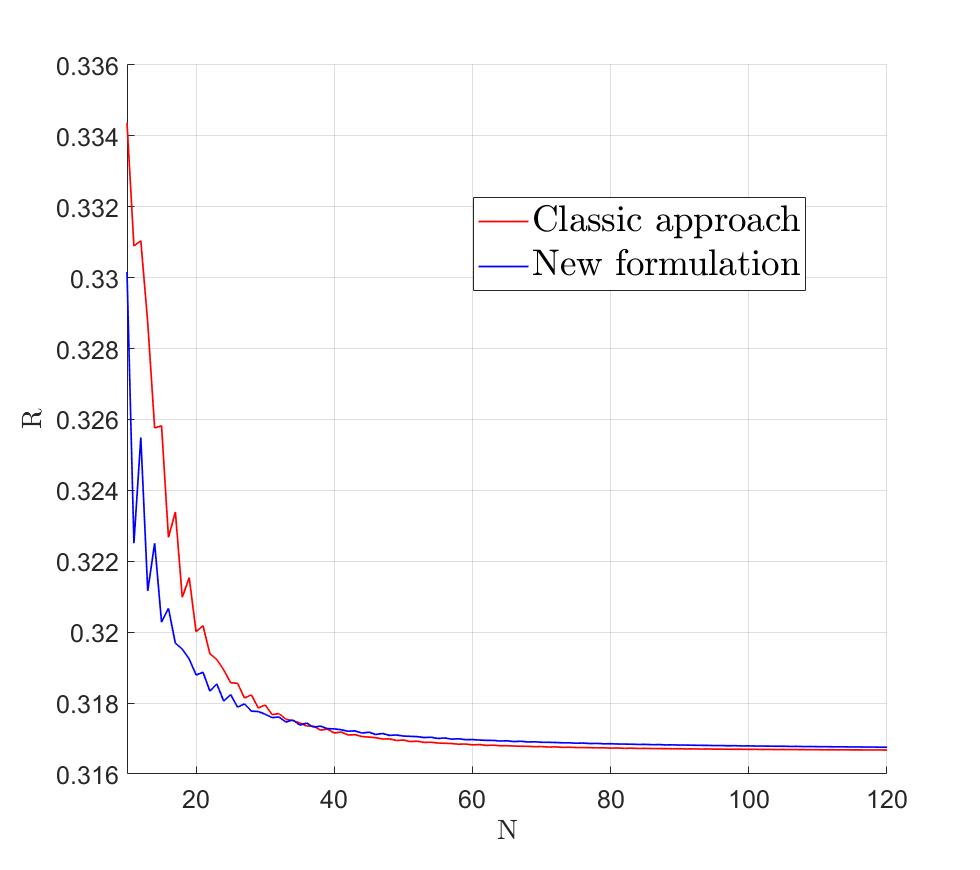}} a) \\
\end{minipage}
\vfill
\begin{minipage}[h]{0.86\linewidth}
\center{\includegraphics[width=1\linewidth]{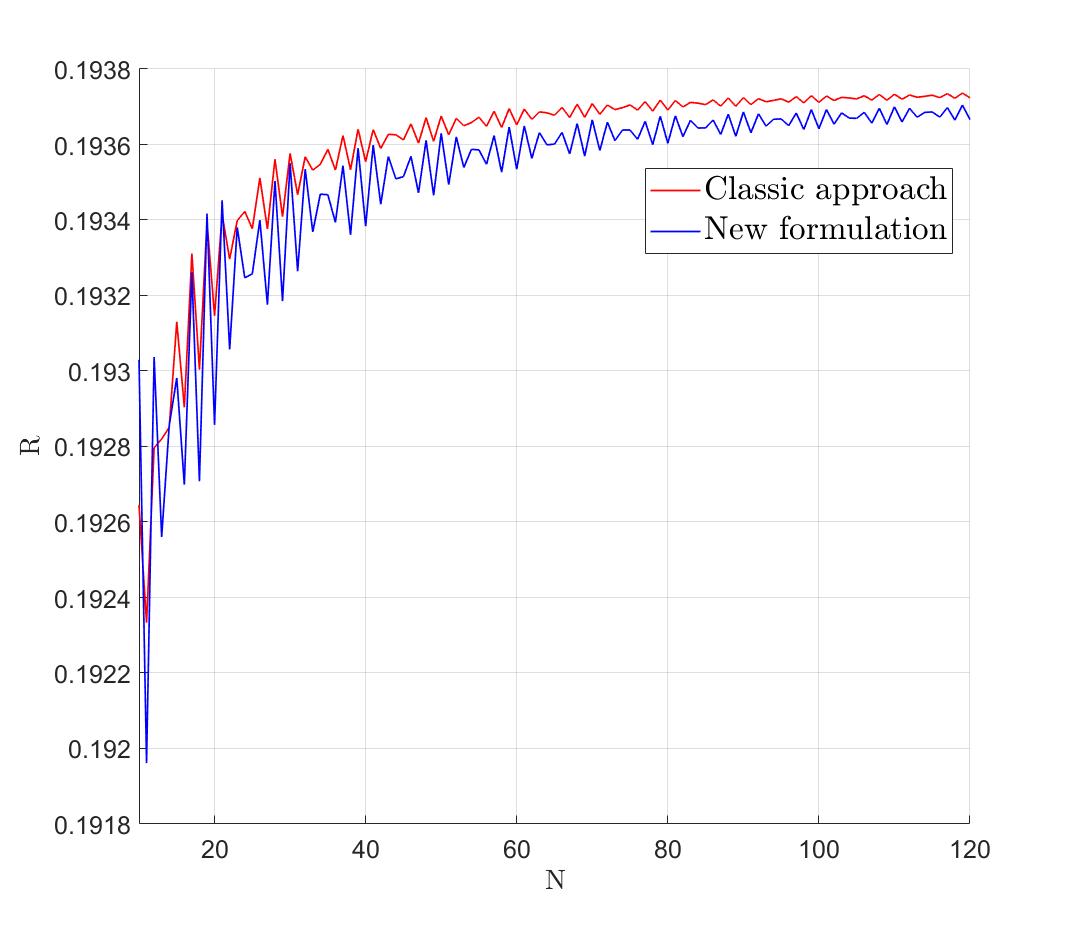}} b) \\
\end{minipage}
\vfill
\begin{minipage}[h]{0.86\linewidth}
\center{\includegraphics[width=1\linewidth]{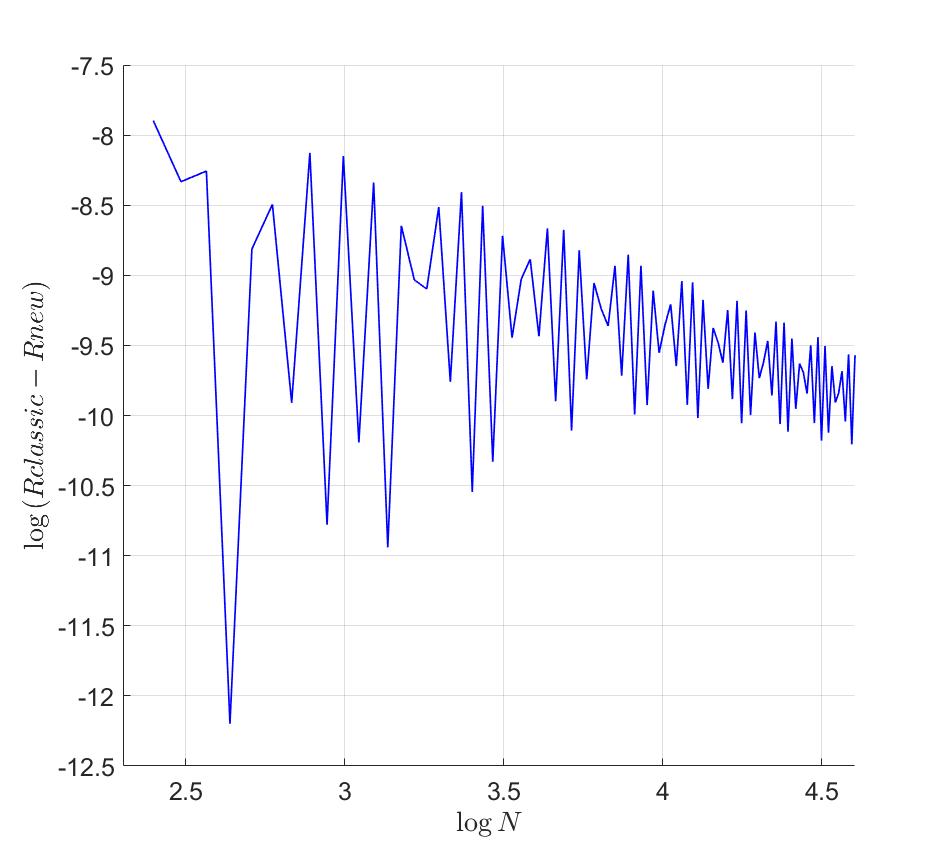}} c) \\
\end{minipage}
\caption{Comparison of the reflection coefficient for the example of a) dielectric grating, b) metal grating of Fig.~\ref{fig:epsf} between the classical method with the Li factorization and  the new approach for increasing number $N$ of the Fourier harmonics, c) difference between the results of the classic and the new approach in case of the metal grating.}
\label{fig:R}
\end{figure}

\begin{figure}[!t]
\begin{minipage}[h]{0.9\linewidth}
\center{\includegraphics[width=1\linewidth]{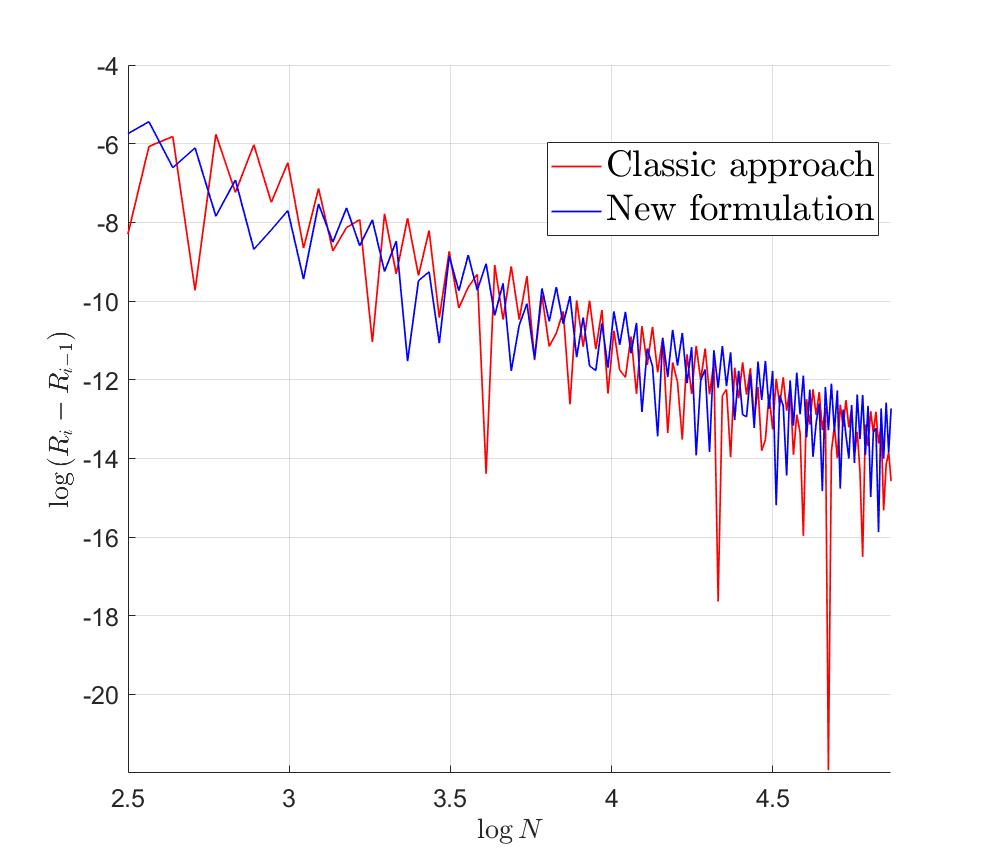}} a) \\
\end{minipage}
\vfill
\begin{minipage}[h]{0.9\linewidth}
\center{\includegraphics[width=1\linewidth]{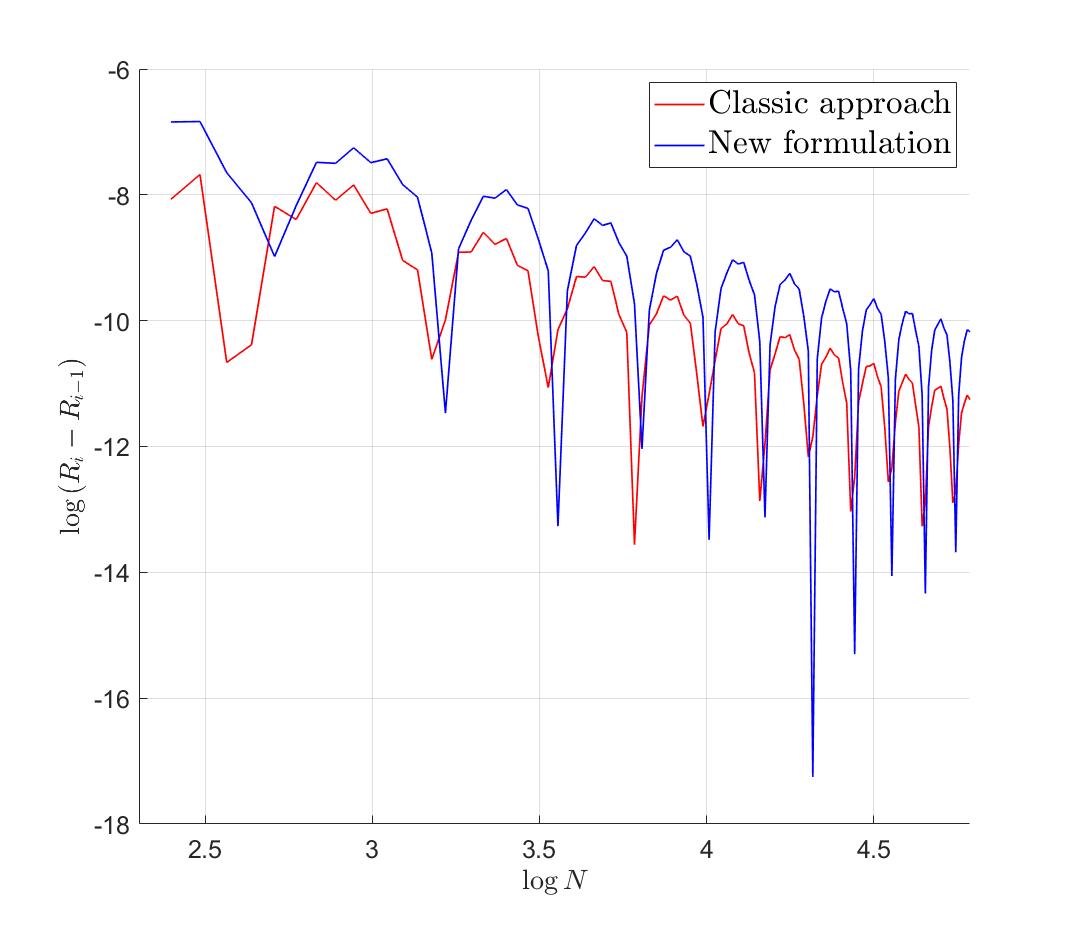}} b) \\
\end{minipage}
\caption{Convergence of the reflection coefficient for the example of a) dielectric grating, b) metal grating of Fig.~\ref{fig:epsf} for the classical method with the Li factorization, and the new approach.}
\label{fig:Rconv}
\end{figure}

Figs.~\ref{fig:R} and \ref{fig:Rconv} demonstrate the relationship between the power reflection coefficient of the slab and its convergence depending on the Fourier series truncation number. Fig.~\ref{fig:R} ensures that both methods give the same answer for the reflection coefficient for both examples, and Fig.~\ref{fig:Rconv} reveals that the rate of the far field convergence is the same for both formulations in each case and is proportional to $1/N^2$.

\begin{figure}[!t]
\begin{minipage}[h]{0.9\linewidth}
\center{\includegraphics[width=1\linewidth]{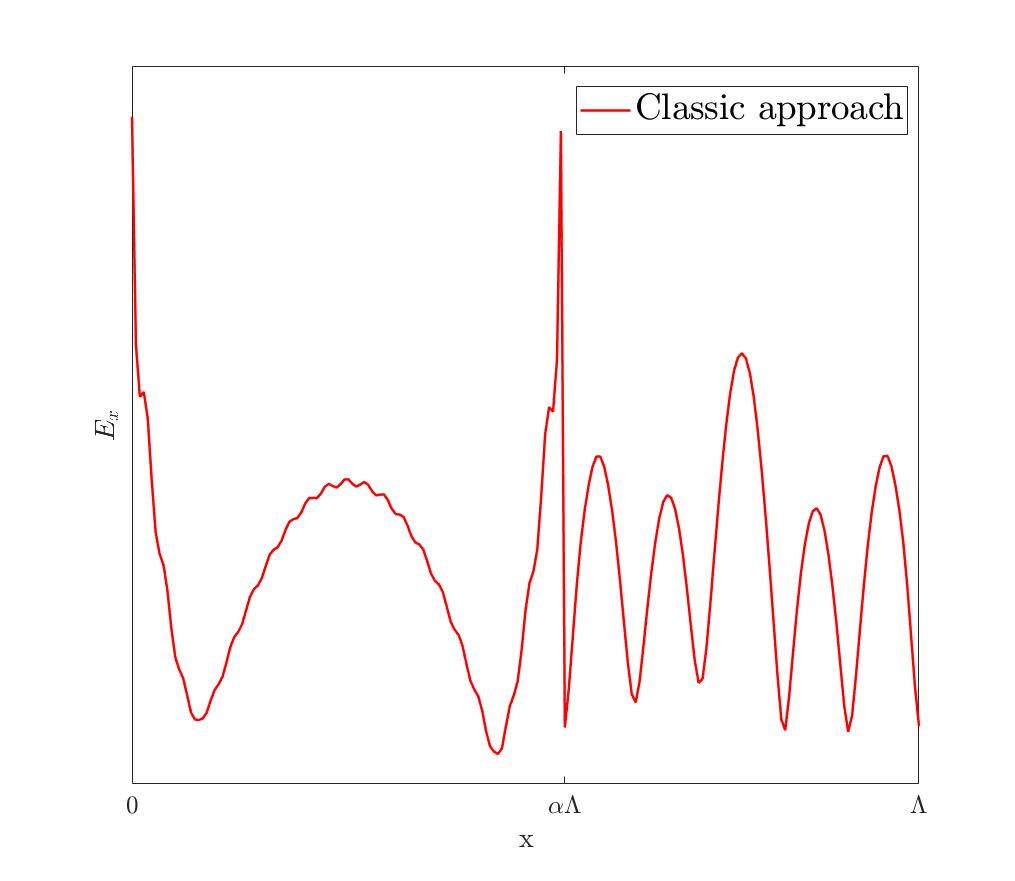}} a) \\
\end{minipage}
\vfill
\begin{minipage}[h]{0.9\linewidth}
\center{\includegraphics[width=1\linewidth]{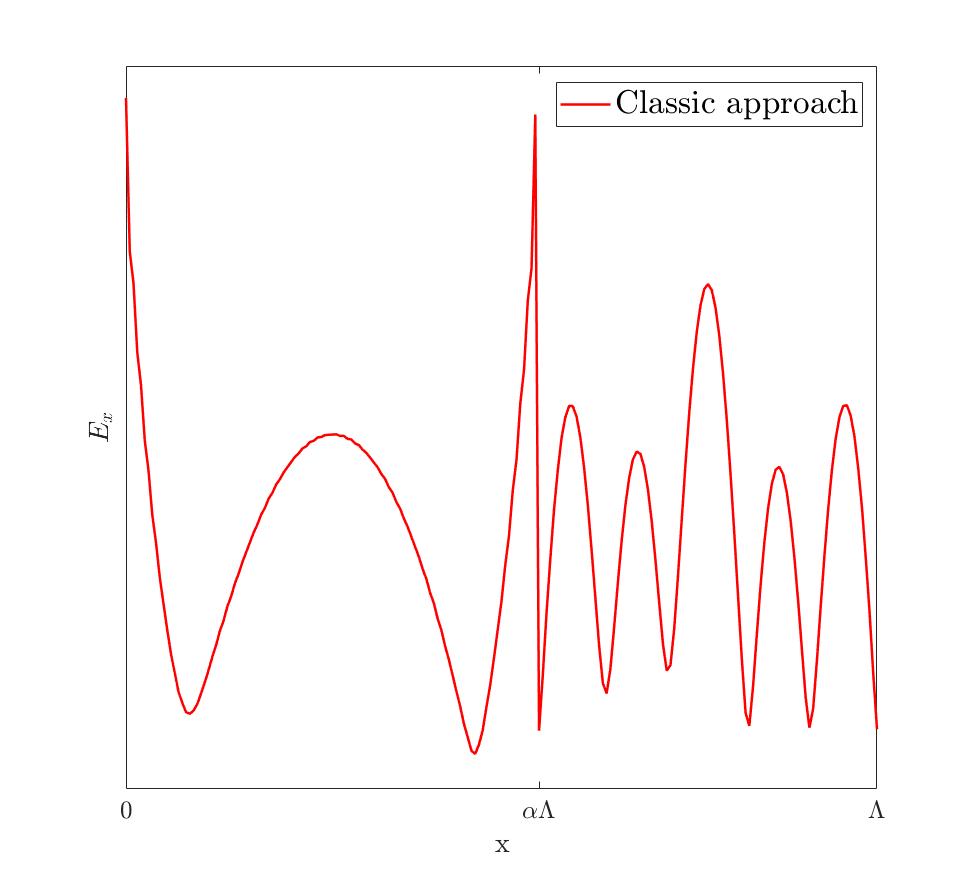}} b) \\
\end{minipage}
\vfill
\begin{minipage}[h]{0.9\linewidth}
\center{\includegraphics[width=1\linewidth]{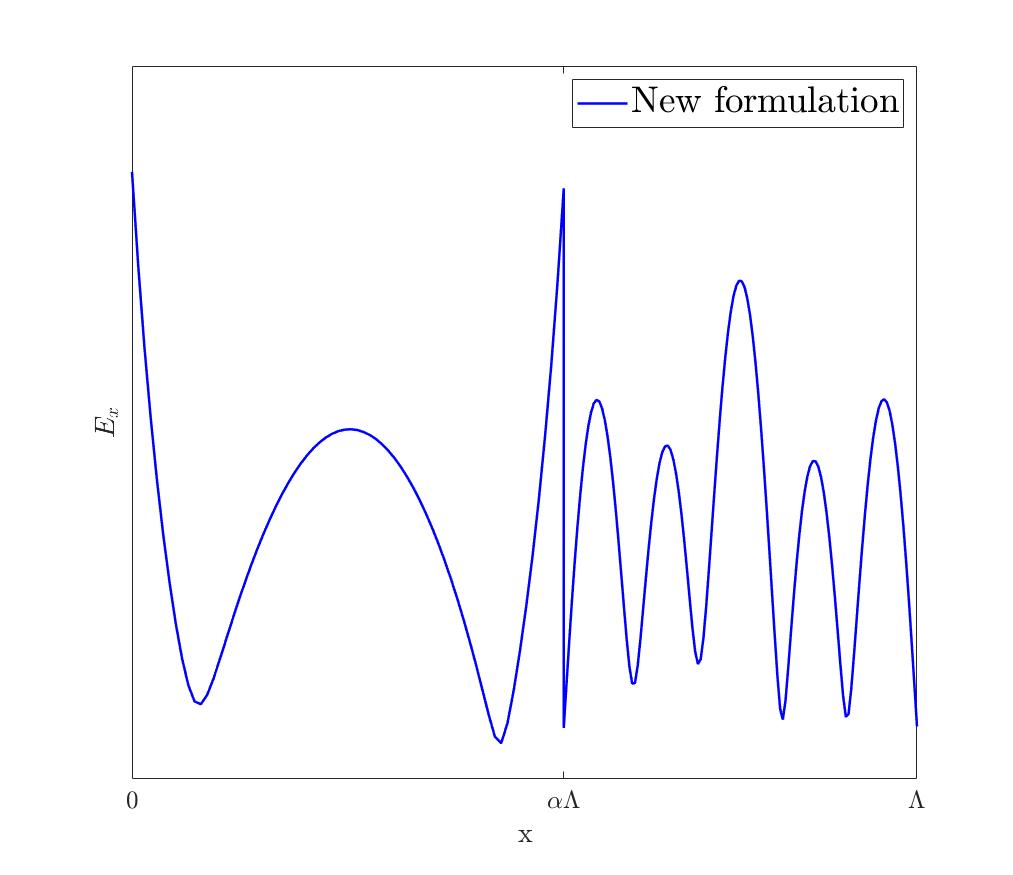}} c) \\
\end{minipage}
\caption{The near field of the dielictric grating example for: a) the classic approach, $N = 40$, b) the classic approach, $N = 90$, c) the new formulation, $N = 40$}
\label{fig:nf}
\end{figure}

\begin{figure}[!t]
\begin{minipage}[h]{0.9\linewidth}
\center{\includegraphics[width=1\linewidth]{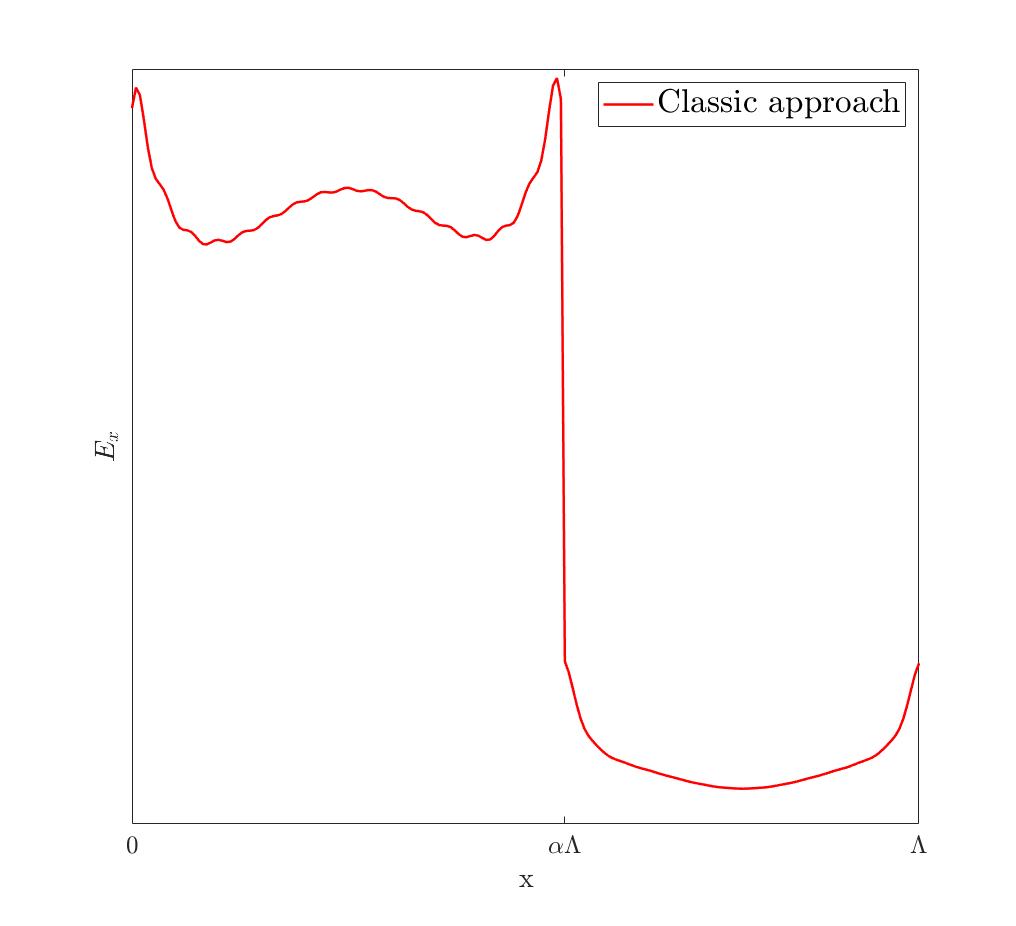}} a) \\
\end{minipage}
\vfill
\begin{minipage}[h]{0.9\linewidth}
\center{\includegraphics[width=1\linewidth]{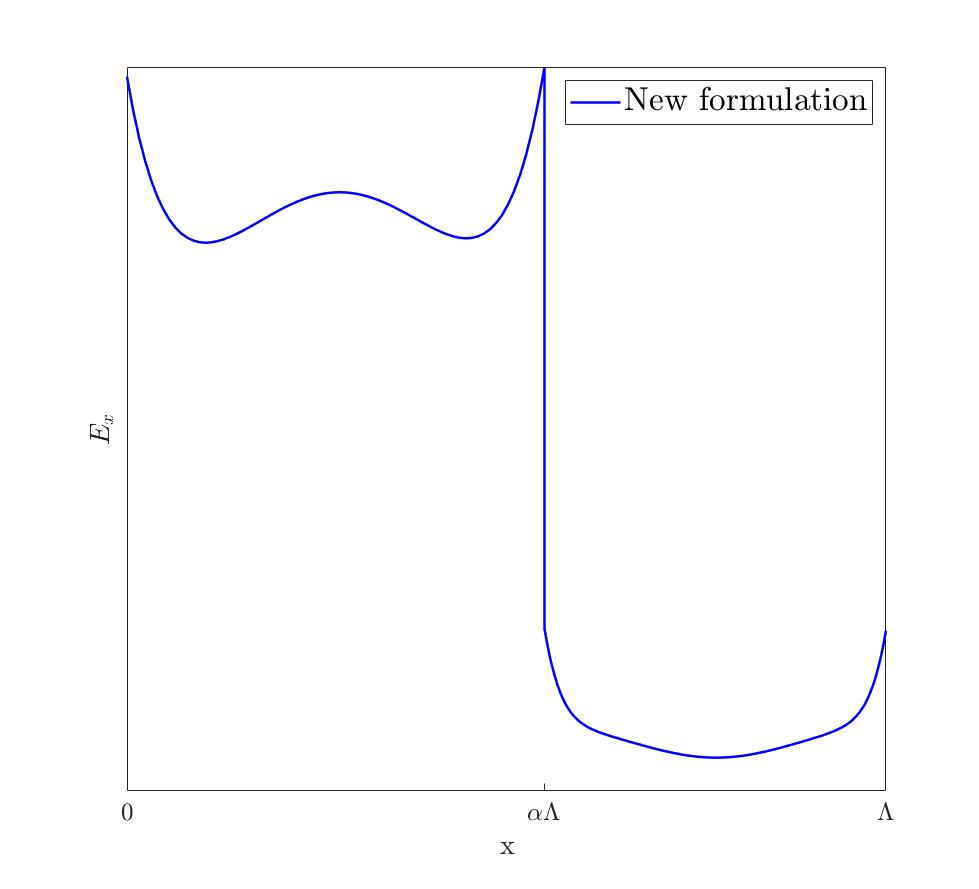}} b) \\
\end{minipage}
\caption{The near field of the metal grating example for: a) the classic approach, $N = 30$, b) the new formulation, $N = 30$}
\label{fig:nf_metal}
\end{figure}

Fig.~\ref{fig:nf} and Fig.~\ref{fig:nf_metal} show the simulation of the near field x-component at z point corresponding to the middle of the structure for two different values of the truncation number $N$ for the classical method with Li factorization rules, and the new approach. To calculate the field in classical approach we use the following equation \cite{lalanne1998computation, weismann2015accurate}:

\begin{equation}
    E_x(x,z)=\dfrac{1}{\varepsilon(x)}\sum_{m \in \mathbb{Z}} \exp{\left(ik_{x_m}x\right)}D_{x_m}(z)
\end{equation}

The field attained from the classic FMM is clearly distorted by oscillations which arise due to the influence of the Gibbs effect. In the case of a large number of Fourier harmonics oscillations accumulate near the interfaces, so the field value may converge quite slowly. The new approach eliminates this distortions. The graphs reveal smooth and correct field functions without oscillations even for relatively small numbers of harmonics. The new method has one more advantage, since an accurate form of the near field is obtained directly from the calculated Fourier amplitude vectors.

\begin{figure}[!t]
\begin{minipage}[h]{0.9\linewidth}
\center{\includegraphics[width=1\linewidth]{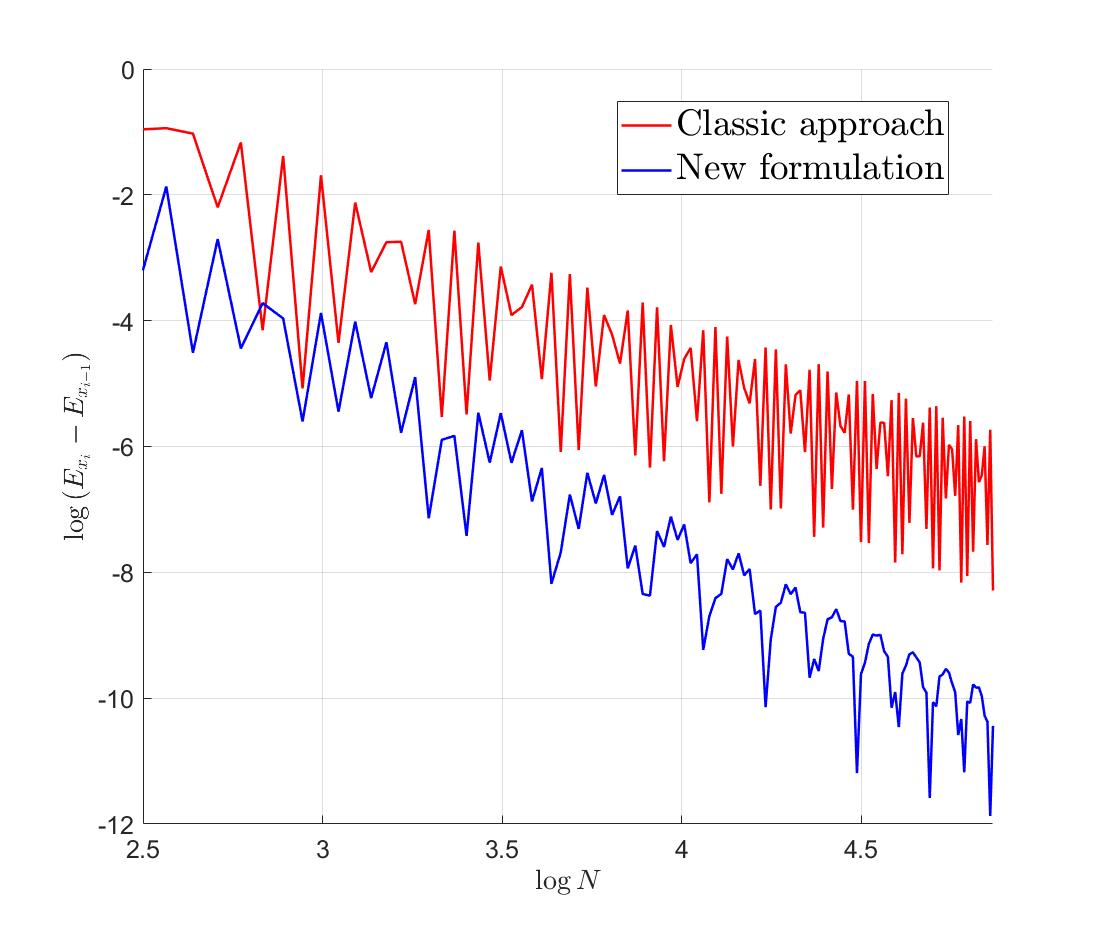}} a) \\
\end{minipage}
\vfill
\begin{minipage}[h]{0.9\linewidth}
\center{\includegraphics[width=1\linewidth]{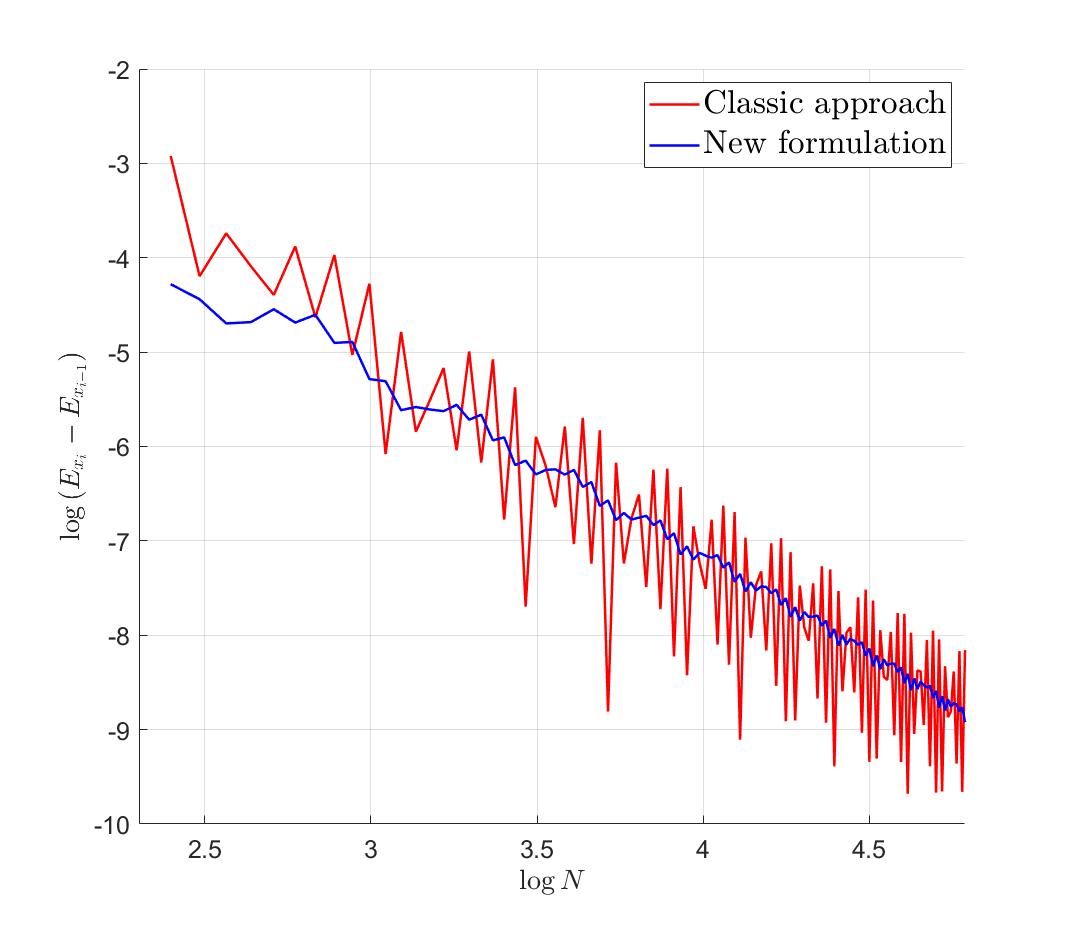}} b) \\
\end{minipage}
\caption{Comparison of the near field convergence for a) the dielectric grating, b) the metal grating in the vinicity of the discontinuity point, $x = 0.5\alpha\Lambda\left(1 - 10^{-3}\right)$, between the classic formulation with the Li's factorization rules and the new approach.}
\label{fig:nfconv}
\end{figure}

Fig.~\ref{fig:nfconv} demonstrates the comparison between the near field convergence attained by the classic and new formulations at a point in the vicinity of the discontinuity. The convergence rate for the new formulation is proportional to $1/N^2$ for both examples, which shows that the convergence of the method has not worsened compared to the classical approach.

\section{Conclusion}
\label{sec:conclusion}
In this paper we proposed a new viewpoint on treatment of field discontinuities at material interfaces within the framework of the FMM. Our approach is based on the explicit consideration of interface conditions for a discontinuous part of the electric field. The new rules for the eigenvalue problem formulation are considered as an alternative to the existing classical method with the Li's factorization rules. The given comparison with the classic method shows that the new approach retains all the advantages of the FMM without inverting the matrix $[[1/\varepsilon]]$, and at the same time it allows highly accurately calculating near field and attaining field pictures free of the Gibbs phenomenon footprint. This advantage comes from the field correction (comparing to the conventional FMM formulation) which is introduced by Eqs.~(\ref{eq:ex_exc}), (\ref{eq:ex_eps_exc}) in the self-consistent manner. This correction appears to keep all the nessessary information about the jump discontinuities as it comes from the extended finite basis where the Fourier harmonics are enriched with appropriate discontinuous functions. Further we plan expand the idea to 2D photonic crystal slabs and we believe that the new method may find its applications in the fields of nonlinear optics and sensing.

\section{Acknowledgements}
\label{sec:acknowledgements}
The work was supported by the Russian Science Foundation, grant No. 22-11-00153.
~\\
\renewcommand{\theequation}{A.\arabic{equation}}
\setcounter{equation}{0}

\section*{Appendix}

In this Appendix we provide details on our implementation of the new method and derivation of the outlined equations. The choice of the functions $g_k(x)$ in Eq.~(\ref{Ed_gk}) is not unique. 
Our choice of these functions is based on the possibility of simple analytical integration of their products with exponential factors, required for derivations of the matrix elements of $Q$ and $P$. Explicitly,
\begin{equation}\label{eq:gfunc}
    \begin{cases}
    n_1 \in \mathbb{Z}, \quad n_2 = n_1 + \alpha \\
    g_{1,2}(x) = -\left(\dfrac{x}{\Lambda} - n_{1,2}\right) +\dfrac{1}{2}, 
    \quad x \in \left[n_{1,2}\Lambda, \left(n_{1,2}+1\right)\Lambda\right] \\
    \end{cases}
\end{equation}
For simplicity we take here $N_d = 2$, though, all derivations can be straightforwardly generalized for an arbitrary number of discontinuities. The functions (\ref{eq:gfunc}) are linear periodic piece-wise continuous functions each having a single discontinuity point on a period, and they are shown in Fig.~\ref{fig:epsf}.

Owing the functions $g_k(x)$ explicitly and accounting for the decomposition of Eq.~(\ref{Ex_cd}), derivation of Eqs.~(\ref{eq:ex_exc}) and (\ref{eq:ex_eps_exc}) is straightforward. The $x$-field Fourier component
\begin{multline}\label{eq:Ex_n}
    \left[E_{x}\right]_{n}=\dfrac{1}{\Lambda} \intop_{0}^{\Lambda} E_{x} \left(x,z\right) \exp\left(-ik_{xn}x\right) = \\ = \left[E_{x}^{c}\right]_{n}+\sum_{m}\mathcal{P}_{nm}E_{xm}^{c}\left(z\right)
\end{multline}
with
\begin{multline}\label{eq:P-matrix}
    \mathcal{P}_{nm} = \sum_{k}\left(\dfrac{1}{\Lambda}\intop_{0}^{\Lambda} g_{k}\left(x\right)\exp\left(-ik_{xn}x\right)\right) \times \\ \times \left(\sum_{q}G_{kq}\exp\left(ik_{xm}x_{q}\right)\right)
\end{multline}
For the product of the field and the permittivity in accordance with Eqs.~(\ref{Ed_gk}) and (\ref{Exd}),
\begin{multline}
    \varepsilon E_{x}(x,z) =\varepsilon E_{x}^{c}\left(x,z\right)+ \\ + \varepsilon\sum_{m}E_{xm}^{c}\left(z\right)\left[\sum_{k}g_{k}\left(x\right)\left(\sum_{q}G_{kq}\exp\left(ik_{xm}x_{q}\right)\right)\right]
\end{multline}
and consequently,
\begin{multline}\label{eq:[epsEx]_n}
    \left[\varepsilon E_{x}\right]_{n}=\dfrac{1}{\Lambda}\intop_{0}^{\Lambda}\varepsilon E_{x}\left(x,z\right)\exp\left(-ik_{xn}x\right) = \\ = \left[\varepsilon E_{x}^{c}\right]_{n}+\sum_{m}Q_{nm}E_{xm}^{c}\left(z\right)
\end{multline}
with
\begin{multline}\label{eq:Q-matrix}
        Q_{nm} =  \sum_{k}\left(\dfrac{1}{\Lambda}\intop_{0}^{\Lambda}\varepsilon\left(x\right)g_{k}\left(x\right)\exp\left(-ik_{xn}x\right)\right) \times \\ \times \left(\sum_{q}G_{kq}\exp\left(ik_{xm}x_{q}\right)\right)
\end{multline}
Integrals in Eqs.~(\ref{eq:P-matrix}) and (\ref{eq:Q-matrix}) with functions (\ref{eq:gfunc}) and piece-wise constant permittivity are found analytically. They read for the matrix $Q$

\begin{multline}
    \dfrac{1}{\Lambda}\intop_{0}^{\Lambda}\varepsilon\left(x\right)g_{1}\left(x\right)\exp\left(-ik_{xn}x\right)  = \\ = \begin{cases} \dfrac{i}{\left(2\pi n\right)}\left[\alpha\left(\varepsilon_1 -\varepsilon_2\right)\left(\dfrac{\sin{\left(\pi m \alpha\right)}}{\pi m \alpha} - \exp{\left(-i\pi m \alpha \right)}\right) - \right. \\ \biggl. - i \varepsilon_1 \sin{\left(\pi m \alpha\right)} - \varepsilon_2 \cos{\left(\pi m \alpha\right)}\biggl], \quad n \neq 0\\
    \dfrac{1}{2}\left(\alpha -\alpha^2\right)\left(\varepsilon_1 - \varepsilon_2\right), \quad n = 0
    \end{cases}
\end{multline}

\begin{multline}
    \dfrac{1}{\Lambda}\intop_{0}^{\Lambda}\varepsilon\left(x\right)g_{2}\left(x\right)\exp\left(-ik_{xn}x\right) = \\ = \begin{cases} \dfrac{i}{\left(2\pi n\right)}\left[\alpha\left(\varepsilon_1 -\varepsilon_2\right)\left(\dfrac{\sin{\left(\pi m \alpha\right)}}{\pi m \alpha} - \exp{\left(i\pi m \alpha \right)}\right) + \right. \\ \biggl. + i \varepsilon_1 \sin{\left(\pi m \alpha\right)} - \varepsilon_2 \cos{\left(\pi m \alpha\right)}\biggl], \quad n \neq 0\\
    \dfrac{1}{2}\left(\alpha -\alpha^2\right)\left(\varepsilon_1 - \varepsilon_2\right), \quad n = 0
    \end{cases}
\end{multline}

Similarly for the matrix $P$ we can obtain

\begin{multline}
    \dfrac{1}{\Lambda}\intop_{0}^{\Lambda}g_{1}\left(x\right)\exp\left(-ik_{xn}x\right) = \\ = 
    \begin{cases}
    0, \quad n = 0 \\
    \left(- i 2 \pi n\right)\dfrac{\exp{\left(i \pi n \alpha\right)}}{\left(2\pi n\right)^2}, \quad n \neq 0 
    \end{cases}
\end{multline}

\begin{multline}
    \dfrac{1}{\Lambda}\intop_{0}^{\Lambda}g_{2}\left(x\right)\exp\left(-ik_{xn}x\right) = \\ = 
    \begin{cases}
        0, \quad n = 0 \\
        \left( - i 2\pi n\right)\dfrac{\exp{\left(-i \pi n \alpha\right)}}{\left(2\pi n\right)^2}, n \neq 0
    \end{cases}
\end{multline}

\printbibliography




\end{document}